\documentclass[aps,prb,preprint,superscriptaddress]{revtex4-1}
\usepackage{ucs}
\usepackage[utf8x]{inputenc}
\usepackage{amsmath}
\usepackage{amsfonts}
\usepackage{array}
\usepackage[francais]{babel}
\usepackage{fontenc}
\usepackage[pdftex]{graphicx}
\usepackage[pdftex]{hyperref}

\usepackage{color}
\definecolor{violet}{rgb}{0.5,0,0.5}
\definecolor{vert}{rgb}{0,0.65,0}

\begin{document}

%%%%%%%%%%%%%%%%%%%%%%%%%%%%%%%%%%%%%%%%%%%%%%%%%%%%%%%%%%%%%%%%%%%%%%%%%%%%%%%%%%%%%%%%%%%%%%%%%%%%%%%%%%%%%%%%%%%%%%%%%%%%%%%%%%%%%%%%%%%%%%%%%%%%%%
%%%                                                                  TEXTE :                                                                  %%%
%%%%%%%%%%%%%%%%%%%%%%%%%%%%%%%%%%%%%%%%%%%%%%%%%%%%%%%%%%%%%%%%%%%%%%%%%%%%%%%%%%%%%%%%%%%%%%%%%%%%%%%%%%%%%%%%%%%%%%%%%%%%%%%%%%%%%%%%%%%%%%%%%%%%%%

\title{Local Optical Probe of Motion and Stress in a multilayer graphene NEMS}

\author{Antoine Reserbat-Plantey}
\author{La\"etitia Marty}
\author{Olivier Arcizet}
\author{Nedjma Bendiab}
\author{Vincent Bouchiat}
\affiliation{Institut N\'eel, CNRS et Universit\'e Joseph Fourier, BP 166, F-38042 Grenoble Cedex 9, France}

\date{\today}
\maketitle

\clearpage
\textbf{Nanoelectromechanical systems (NEMSs)\cite{reviewNEMS} are emerging nanoscale elements at the crossroads between mechanics, optics and electronics, with significant potential for actuation and sensing applications.
The reduction of dimensions compared to their micronic counterparts brings new effects including sensitivity to very low mass\cite{{bachtoldbalance},{zettlbalance}}, resonant frequencies in the radiofrequency range \cite{zettlNTRF}, mechanical non-linearities\cite{BachtoldNonlinear} and observation of quantum mechanical effects\cite{quantumNEMS}.
An important issue of NEMS is the understanding of fundamental physical properties conditioning dissipation mechanisms, known to limit mechanical quality factors and to induce aging due to material degradation. 
There is a need for detection methods tailored for these systems which allow probing motion and stress at the nanometer scale.
Here, we show a non-invasive local optical probe for the quantitative measurement of motion and stress within a multilayer graphene NEMS provided by a combination of Fizeau interferences, Raman spectroscopy and electrostatically actuated mirror. 
Interferometry provides a calibrated measurement of the motion, resulting from an actuation ranging from a quasi-static load up to the mechanical resonance while Raman spectroscopy allows a purely spectral detection of mechanical resonance at the nanoscale.
Such spectroscopic detection reveals the coupling between a strained nano-resonator and the energy of an inelastically scattered photon, and thus offers a new approach for optomechanics.}

%%%%%%%%%%%%%DEBUT DE L'INTRO%%%%%%%%%%%%%%%%%%%%
Graphene's\cite{Novoselov2004p2970} outstanding mechanical \cite{HoneAFM2008}, electrical \cite{castroneto2009} and optical \cite{Nair2008} properties, make it an ideal material for flexible, conductive and semi-transparent films.
Multilayer graphene (MLG), which has a thickness of several tens of atomic layers, is sufficiently stiff \cite{Booth2008} to produce free-standing cantilevers, with an unprecedented aspect ratio.
Such structures can be used to make suspended mirrors, with a mass ranging from tens to hundreds of femtograms. 
When suspended over silica, such cantilevers form optical cavities which can be electrostatically actuated and, are thus ideal for the implementation of NEMS\cite{Bunch}.
Previous attempts to probe local motion of graphene resonators\cite{BachtoldAFM} have reached nanometer scale but cannot measure directly the stress and remained confined to a limited range of pressure and temperature.
Nevertheless, recent studies\cite{Bolotin} based on hybrid graphene-metallic cantilevers has brought promising results on static stress graphene using optical profilometry. 
In the present work, we use Raman spectroscopy to probe the local stress within a MLG cantilever.
We explore mechanical regimes from DC up to MHz frequencies by taking advantage of the large dynamical range of optical detection.
The MLG displacement is considerably greater than previously reported, and optical interferences allow self-calibration of displacements, while Raman spectroscopy gives quantitative analysis of the local stress within the structure.

\bigskip
%%%%%%%%%%%%%%%%%FIGURE 1 : Fizeau fringes in a MLG cantilever overhanging silicon oxide.%%%%%%%%%%%%%%%%%%%%%%%%%%%%
\begin{figure}[htbp]
\begin{center}
	\includegraphics[width=15cm]{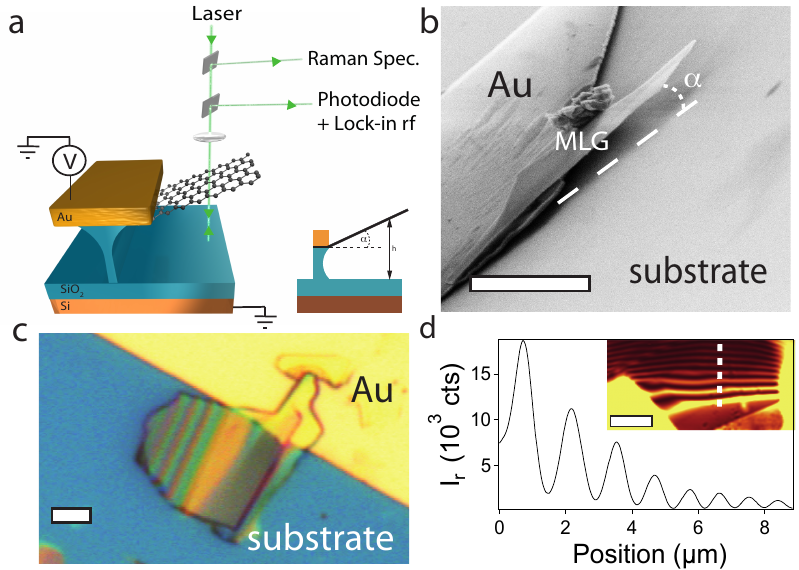}
	     \caption{\textbf{Fizeau fringes in a MLG cantilever overhanging silicon oxide.}  \textbf{a}: typical SEM micrograph of a sample, consisting in two reflectors: an oxidized silicon back-mirror and a MLG cantilever, having a semi-transparent behavior. \textbf{b}: Schematic view of the device : electrical excitation and optical detection, either with a photodiode (intensity) or a Raman spectrometer (intensity and spectral data). \textbf{c}: White light optical image of a device showing iridescence. \textbf{d}: Reflectance profile is measured along the dashed-line of the inset. The reduction in signal strength observed at large distances from the hinge is due to reduced spatial mode matching. The fringe contrast is however preserved. Inset: Reflectance confocal (X,Y) scan at 633 nm. The scale length is 5 $\mu$m.}\label{fig:fig1}
\end{center}
\end{figure}

Samples were prepared from micron-sized MLG planar flakes clamped on one side by a gold film and overhanging silicon oxide (see Methods).
Typical samples had a thickness of approximately 100 monolayers (\textit{ca}. 30 nm), as verified by atomic force microscopy (cf Supp. Info.).
Their thicknesses were adjusted so as to prevent collapse whilst maintaining semi-transparency with an optical reflectance (transmission) coefficient  $R$ = 0.22 ($T$ = 0.61) for a 30 nm thick MLG\cite{Skulason}.
Some flakes tend to stick up after the fabrication process (see Fig. \ref{fig:fig1}a), at a wedge angle $\alpha \rm{\in [5^{\circ};35^{\circ}]}$, and these leave a wedge gap of length $h(x,y)$ in the range between 0.3 and 3 microns.
The resulting structures form an optical cavity, characterized by a low optical interference order ($2h/\lambda$<10, where $\lambda$ is the probe wavelength), with MLG top mirror of extremely low mass (10-100 fg), and of high mechanical resonance frequencies (1-100 MHz).
With approximately 100 measured samples, we have observed a variety of geometries (Fig. \ref{fig:fig1}), allowing us to explore various mechanical regimes, with a wide range of wedge angles $\alpha$, sizes, and shapes.
Iridescence is observed under white light illumination (Fig. \ref{fig:fig1}c), and the interference pattern observed under monochromatic light (Fig. \ref{fig:fig1}d) has contrasted equal-thickness fringes (so-called Fizeau fringes, see Supp. Info.).
Unlike conventional graphene-based optical cavities with fixed geometries\cite{Ling2010}, the optical length of the cavity increases linearly along the cantilever, which allows the observation of multiple interference fringes (cf. Fig. \ref{fig:fig1}cd).
Interference patterns are observed both from the reflection of the pump laser and from Raman scattered light (see Supp. Info.), the latter having the considerable advantage of carrying local informations related to the material (stress, doping, defects, temperature). 

Furthermore, the optical length of the cavity can be adjusted through electrostatic actuation of the cantilever, thus producing rigid shift of the interference fringes pattern (see video in Supp. Info.).
This is achieved by applying a DC or AC voltage $V$ (typically up to $30$ Volts) to the clamp electrode (Fig. \ref{fig:fig1}b) while the SiO$_2$ capped silicon substrate is grounded \cite{Bunch}. 
This results in an attractive electrostatic force $F$, which produces reduction of the cavity length with respect to the equilibrium position $h_0$ in absence of driving.
We measure the response of a harmonic drive, which create of force quadratic in voltage $F(2\omega) \propto V(\omega)^2$ through the local light intensity variation, $\Delta I(x,y,2\omega,h_0)$: 
\begin{equation}
\Delta I(x,y,2\omega,h_0) \propto \chi_{mec} (x,y,2\omega) \chi_{opt} (x,y,h_0) \ V(\omega)^2 ,
\label{reso}
 \end{equation}
where $\chi_{mec}$ is the mechanical susceptibility (see Supp. Info.) and  $\chi_{opt}$ is the optical susceptibility defined as $\chi_{opt} (x,y,h_0) = \partial g_{opt} / \partial h$, where $g_{opt}$ is a periodic interferometric function of $h(x,y)$ defined as the normalized reflected light ($I_r$) or Raman scattered light ($I_G$) $I_{r,G}/I_0 = g_{opt}(h)$ (see Fig. \ref{fig:fig1}d).
\bigskip
%%%%%%%%%%%%%%%%%%%%%%%%%%Stress mapping and quasi static actuation of wedged MLG NEMS.%%%%%%
\begin{figure}[htbp]
\begin{center}
	\includegraphics[width=15cm]{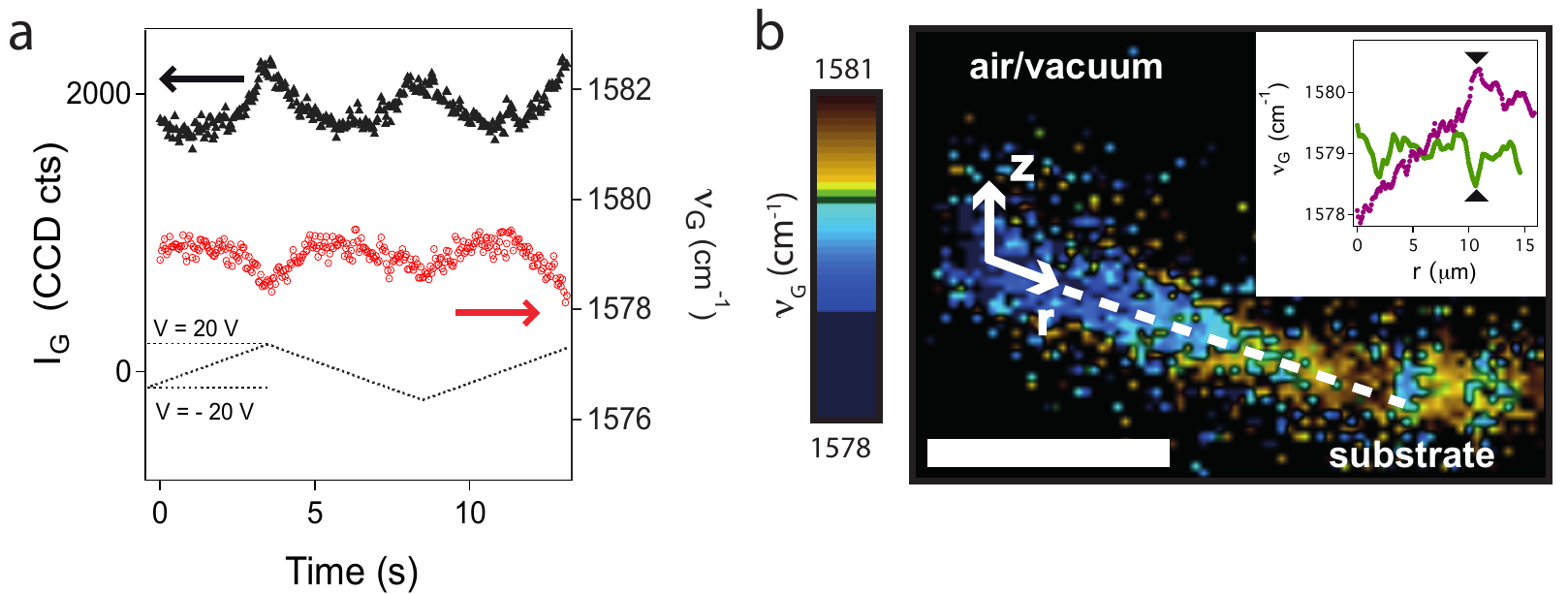}
	     \caption{\textbf{Quasi static actuation and stress mapping of wedged MLG NEMS.}  \textbf{a}: Variations in G peak intensity (black) and position  (red) under MLG actuation, revealing peak softening. The lower dashed line represents the drive voltage. \textbf{b}:  Map of G peak produced by confocal (X,Z) scan mapping of the cantilever cross section. Inset:  G peak position along the cantilever (purple, and along the same sample following collapse of the cantilever onto the silica substrate (green). Black marks indicate the hinge position. The scale bars represent a length of 5 $\rm{\mu}$m.}\label{fig:fig1bis}
\end{center}
\end{figure}

The quadratic dependence of $\Delta I$ upon voltage is systematically observed, both for reflected light (cf. Fig. S5) and for the MLG Raman lines (cf. Fig. \ref{fig:fig1bis}a).
Since $g_{opt}$ is $\lambda/2$ periodic, a precise calibration of the low frequency motion response under electrostatic actuation can be obtained and is found to be of the order of 20 nm.V$^{-2}$. (cf. Supp. Info.).
Interestingly, energy of the stress-sensitive optical phonon (so called Raman G peak) also follows quadratic behavior.
The G peak Raman shift is indeed synchronized with the interferential response $I_G(t)$ (Fig. \ref{fig:fig1bis}a), and exhibits softening of approximately 1.9 cm$\rm{^{-1}}$ at the maximum cantilever deflection.
This Raman peak softening cannot be interpreted as a doping effect since the doping level necessary to induce the observed Raman shifts would correspond to surface charge incompatible with the one induced by the gate drive\cite{Kim}. 
Moreover, the doping induced during AC gating would directly follow gate variation and therefore be $\omega$ periodic which is in disagreement with the observed $2\omega$ periodic Raman shift (cf. Fig. \ref{fig:fig1bis}a).
This Raman peak softening is interpreted as a stress/strain effect and, by analogy with strained graphene measurements\cite{{Otakar2010},{Otakar2011},{Hone2009}}, it is thus possible to extract a corresponding strain value of $0.06\%$ at maximum deviation resulting from a quasi static stress of 600 MPa.
For such low strain, the G band splitting is not resolved.
Besides, the stress exerted at the hinge scales like $LF/[2t^2\rm{sin}(2\alpha)] $ where $F$, $L$ and $t$ are the electrostatic force, the cantilever length and thickness, respectively (see Supp. Info.).
For our large aspect ratio structures ($L/t\gg1$), the local stress can be very intense and reaches hundreds of MPa for electrostatic forces estimated here, which are about 25 nN.$\mu$m$^{-1}$. 
This value is in agreement with the quasi-static stress of 600 MPa deduced above.

Nevertheless, the MLG Raman signature depends on the position along the flake.
A micro-Raman confocal (X,Z) scan (Fig. \ref{fig:fig1bis}b) reveals a linear increase in the position of the G peak along the cantilever axis, from the free-end of the cantilever to the hinge, which is not observed when the MLG is collapsed (inset of Fig. \ref{fig:fig1bis}b).
This linear shift could be interpreted as a continuously increasing electrostatic field effect\cite{{Kim},{Ferrari}}, due to charge within the substrate which also influences the position of the Raman G peak\cite{Berciaud2009}.
However, in this experiment, the G peak shows local hardening around the hinge position, in the suspended case, whereas local softening is observed at the same location, after collapse.
Indeed, uniaxial strain in MLG also induces symmetry breaking of the Raman G peak, leading to a mode splitting and, each component (G$^+$, G$^-$) softens or hardens under tensile or compressive strain, respectively\cite{{Otakar2010},{HoneAFM2008}}.
That stress induced Raman shift is characterized by an average mode shift rate about\cite{Otakar2011} -3.2 cm$^{-1}$.GPa$^{-1}$.
This outcome is in agreement with a maximum compressive strain at the hinge in the suspended case and, a transition toward a tensile strain when it collapses.
By converting these Raman shifts into stress at the hinge, this gives an equivalent built-in stress of 300 MPa.
\bigskip

%%%%%%%%%%%%%%%%%%%%%%%FIGURE 2: Detection of mechanical resonance by Fizeau interferometry.%%%%%%%%%%%%%%%%%%%%%%%%%%%
\begin{figure}[htbp]
\begin{center}
		\includegraphics[width=15cm]{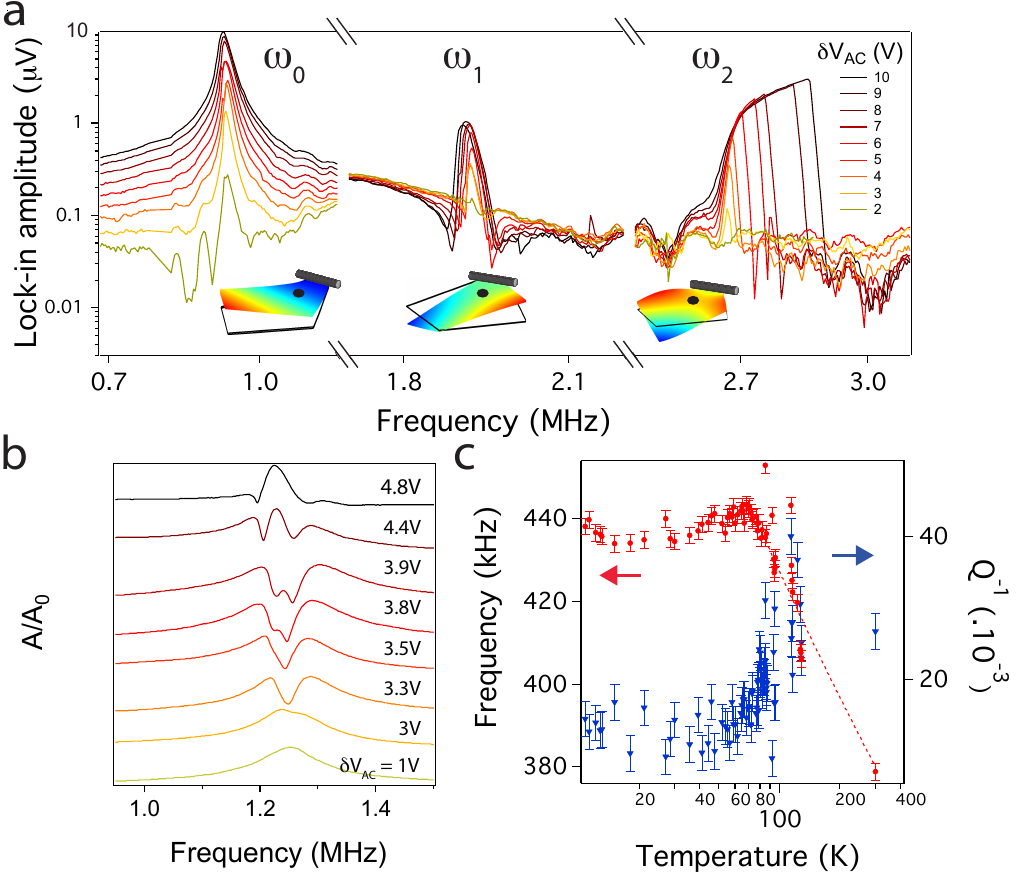}
	     \caption{\textbf{Detection of mechanical resonance by Fizeau interferometry.} \textbf{a}: Amplitude at $2\omega$ versus drive frequency for different rf drive voltages showing the non-linear behavior of the fundamental and the two first harmonic mechanical modes. Laser probe is focused close to the hinge. Schematics of the deformed shape are associated to each mode and dark dots represent the laser probe position. \textbf{b}: Amplitude at $2\omega$ versus drive frequency, for an increasing range of drive voltages (bottom to top), revealing signal folding due to optical interferences. The laser spot is located close to the free end of the cantilever. \textbf{c}: Evolution of MLG cantilever resonance frequency $\omega_0$ (red) and its associated dissipation $Q^{-1}$(blue) as a function of the temperature of the optical cryostat. Measurements a-c are performed under vacuum and the laser spot is positioned at the edge of a fringe ($\chi_{opt}$ is optimal).}\label{fig:fig3}
\end{center}
\end{figure}

Interestingly, changing both the laser spot position and the drive amplitude allows probing in a separated fashion the non-linearities arising from mechanical (Fig. \ref{fig:fig3}a) and from optical (Fig. \ref{fig:fig3}b) origins.  
It is worth noting that optical non-linearities are observed when the probe is far from the hinge (see Fig. \ref{fig:fig3}b) where, due to the lever-arm effect, oscillation amplitude of $h(x,y)$ becomes comparable to the probe wavelength.
Like in quasi-static regime, it is possible to calibrate the displacement amplitude with respect to the driving excitation $\delta V_{AC}$ by using the periodic nature of $\chi_{opt}$.
Peak folding in indeed observed for increasing drive beyond 3V (cf. Fig. \ref{fig:fig3}b). 
Assuming mechanical linear response, the drive increase to produce two successive foldings (corresponding to $\lambda/4$ in amplitude) provides calibration of the drive efficiency, which equals 150 nm.V$\rm{^{-2}}$ in the case presented in Fig. \ref{fig:fig3}b.
Interestingly, the entire signature of the optical non-linearities is visible for a restricted range of drive voltage which ensures to neglect mechanical non-linearities.

Close to the hinge, optical non-linearities are extinguished due to smaller variations of $h$ and thus reveal non-linearities of mechanical origin, which are observed on each mode for drive voltages higher than 4V (cf. Fig. \ref{fig:fig3}a). 
This measurement highlights the wide range of mechanical non-linearities observed in MLG structures\cite{{Lifshitz},{BachtoldNonlinear},{Landau}}, and it is worth noting that the detection efficiency strongly depends on the mode profile since it is based on Fizeau fringes pattern modulation.
$\chi_{mac}(x,y,2\omega)$ can exhibit important variations due to the spatial nature of the probed vibration. 
In particular, $\chi_{mec}(x,y,2\omega)$ can be strongly reduced when the laser probe is focused at a node of the mechanical resonance.
As an example, the first harmonic ($\omega_1$), found to be a torsional mode \textit{via} finite elements analysis, implies cantilever deformation with singular position where the cavity length does not vary (typically, a node region ($x_n, y_n$)).
Thus, according to Eq. \ref{reso}, $\chi_{mec}(x_n,y_n,2\omega_1) \ll \chi_{mec}(x_n,y_n,2\omega_0)$, whereas focusing the laser at a different position allows to enhance the local optical response. 
This particular extinction feature of the detection has a great potential for further mapping of MLG deformation associated with a single mechanical mode.

In order to investigate the influence of the laser probe in our all-optical method, cryogenic measurements has been carried out as shown in Fig. \ref{fig:fig3}c.
The fundamental resonant frequency exhibits a linear upshift upon cooling from 300K to 70K, below which it saturates due to extrinsic heating (see Supp. Info.). 
In contrast to doubly clamped graphene-based NEMS\cite{{Bunch},{BachtoldNonlinear},{Mandar}}, it is not possible to discuss the frequency hardening observed in Fig. \ref{fig:fig3}c in terms of cantilever tensioning induced by differential thermal expansion since we study a simply clamped geometry.
An important feature of any resonator is the measurement of the quality factor, defined as $Q=\omega/\Delta \omega$, which characterizes the high sensitivity (high $Q$) of the resonator to its environment. 
A linear decrease of the dissipation, $Q^{-1}$, is observed upon cooling to 70K.
Both effects, frequency hardening and decrease of the dissipation, are possibly a consequence of the stiffening of the clamp electrode. 
Further measurements will allow to investigate both extrinsic effects (clamp stiffening losses) and mechanical intrinsic properties of MLG which should bring new insights to understand damping mechanisms in NEMS. 
Effective substrate temperature is obtained by measuring Stokes and Anti-Stokes Raman intensities ratio (Supp. Info.) and indicates a temperature threshold of 70K.
Interestingly, all the physical quantities (resonant frequencies, quality factor $Q$, Raman shift) are sensitive to the environmental temperature until 70K. 
This demonstrates experimentally that room temperature experiments discussed in this letter are not altered by laser heating.
Concerning absorption of mechanical energy at resonance, we have seen no change in the Raman Stokes/Anti-Stokes measurements when sweeping the excitation frequency through the mechanical resonance, indicating no increase of phonon bath temperature. 
\bigskip

%%%%%%%%%%%%%%%%%%%%FIGURE 3: Detection of mechanical resonance by Raman spectroscopy%%%%%%%%%%%
\begin{figure}[htbp]
\begin{center}
	\includegraphics[width=15cm]{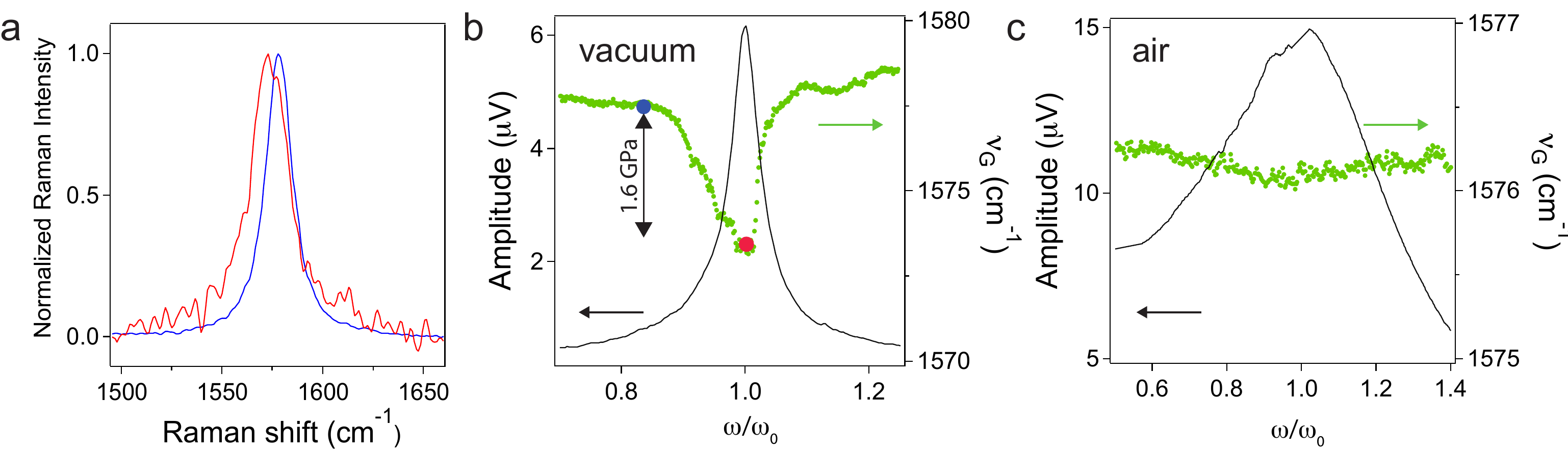}
	     \caption{\textbf{Detection of mechanical resonance and dynamic stress using Raman spectroscopy.} \textbf{a}: Raman spectra of the G peak under MLG actuation at mechanical resonance frequency (red) and off resonance (detuned by 360 kHz) (blue). For the resonating case, signal to noise ratio is smaller than off-resonance case due to larger oscillation amplitude at resonance which takes the resonator out of focus. \textbf{b-c}: Lock-in amplitude at $2\omega$ (dark line) as a function of drive frequency, for a 5V rf drive voltage under vacuum  (\textbf{b}) and in air (\textbf{c}). The position of the Raman G peak (green) shows a softening which coincides with the mechanical resonance. This softening is even more marked under vacuum. Blue and red circles, shown in caption b, correspond to the Raman spectra plotted in Fig. 3a. This sample is the same as presented in Fig. 3a.}\label{fig:fig5}
\end{center}
\end{figure}

To demonstrate the spectral detection of mechanical resonance, Raman response of MLG cantilever under vacuum is plotted (Fig. \ref{fig:fig5}a) at fundamental mechanical resonance $\omega_0$ = 1.2 MHz (red curve) and off resonance (blue curve).
At $\omega_0$, G peak softening is -5 cm$\rm{^{-1}}$ in position and about +10 cm$\rm{^{-1}}$ in width (peak's FWHM) (Fig. \ref{fig:fig5}a) which takes into account the averaging induced broadening (see Supp. Info.). 
This Raman softening estimated at -1 cm$\rm{^{-1}}$.V$\rm{^{-2}}$, is attributed to corresponding variation in stress within the cantilever, induced at mechanical resonance according to universal stress behavior in sp$^2$ carbon materials\cite{{Otakar2011},{Hone2009},{Ferrari2009}} (shift rate: 0.003 cm$^{-1}$.MPa$^{-1}$).
This dynamical stress is thus about 1.6 GPa, and therefore provides a quantitative means of detecting NEMS resonances stress effects.
It is worth noting that the measured stress in MLG cantilever at mechanical resonance is more than one order of magnitude higher than previously reported\cite{Pomeroy} in silicon-based MEMS devices.
 
In Figs. \ref{fig:fig5}b-c, we have detected the fundamental mechanical resonance of this MLG cantilever using both reflected and Raman scattered light under different experimental conditions.
As the lifetime of optical phonons is much shorter (1 ps) than $\omega_0^{-1}$ ($\rm{\sim \ 100 \ ns}$), the Raman scattering process provides instantaneous information related to stress in the vibrating cantilever.
For each excitation frequency, we record a Raman spectrum (1s averaging), which reflects stress at the cantilever position.
For several samples, we were able to check that this softening behavior (green curve, Fig. \ref{fig:fig5}b-c), observed under mechanical excitation, coincides with the mechanical resonance width irrespective of the chamber pressure (cf. Fig. \ref{fig:fig5}b-c).

In contrast to the vacuum case (Fig. \ref{fig:fig5}b), where the quality factor is about $Q_{vac}$ = 26.1, the same sample in air (Fig. \ref{fig:fig5}c) has a reduced quality factor ($Q_{air}$ = 2.3) as well as a shifted Raman G peak, which indicates that the dynamical stress depends on the oscillation amplitude (as also suggested by the Fig. S9 in Supp. Info.).
The value of $Q_{air}$ agrees with typical viscous damping model\cite{Hosaka} for that particular geometry, thus confirming that viscous damping is the predominant mechanism for limiting the quality factor in air (see Supp. Info.).
Nevertheless, this damping mechanism is no longer the main one under vacuum where the dissipation may be governed by clamping losses.
One can compare the ratio of the drive efficiency at low frequency (20 nm.V$\rm{^{-2}}$) and at resonance (150 nm.V$\rm{^{-2}}$) with the ratio of the G peak shift sensitivity at low frequency (1 cm$\rm{^{-1}}$.V$\rm{^{-2}}$) and at resonance (5 cm$\rm{^{-1}}$.V$\rm{^{-2}}$).
Both ratios equal to the measured quality factor $Q$ ($\rm{\sim 6}$), as expected for a mechanical resonator\cite{Landau}.
To demonstrate the versatility of the Raman-based spectral detection of the mechanical resonance, we have investigated a similar effect on two other types of NEMS (Si nano cantilevers and SiC nanowires, see Supp. Info.).

There exists a fundamental coupling between the device position (directly given by the cavity length $h(x,y)$) and a spectroscopic property (measured by the Raman peak shift). 
In our case, the flake displacement generates a mechanical stress that causes a shift of stress-sensitive Raman peaks. 
We therefore extract a coupling constant that is the ratio between the Raman peak shift (in Hz) and the estimated displacement (in meters). 
The novel optomechanical coupling linking the Raman G peak position shift to the cantilever displacement reaches $\rm{\sim10^{17}}$ Hz.m$\rm{^{-1}}$ in magnitude, which compares favorably to similar quantities involving other optomechanical systems\cite{PRL arcizet 2006}.
This large optomechanical coupling, in which all the isotropically scattered Raman photons carry informations on the nano-resonator dynamics, enables mechanical stress information to be spectrally encoded.
Interestingly, this provides an efficient rejection of background signal even in a backscattering configuration for on-chip devices.
For the detection of submicron NEMS, this generates in many cases better signal to noise ratio, compared to diffraction-limited elastic optical detection techniques.
Finally, the resonant nature of Raman scattering in graphene preserves a large interaction cross-section, allowing the optomechanical coupling to be maintained even when working with nanosized oscillators, which is not the case in standard optomechanical approaches\cite{{PRL arcizet 2006}} in which only a small fraction of the detected photons carries the optomechanical information.

\bigskip

%%%%%%%%%%%%%%%%PERSPECTIVES + CONCLUSION %%%%%%%%%%%%%%%%%%%%%%%%%

In conclusion, we demonstrate a non invasive and high bandwidth optical probe, enabling imaging of dynamical stress and motion in a NEMS.
This probe, combining Raman spectroscopy with Fizeau interferometry, is applied to multilayer graphene NEMS and is found to be compatible with two other types of NEMSs.
Calibrated motion and stress can be measured and mechanical resonances can be detected through optical mode shifting and mapped as a local stress along a vibrating cantilever. 
The reliability of Raman spectroscopy in this context finds its origins in the large optomechanical coupling between strain modulation and mechanical displacements.
We demonstrate the coupling between flexural vibrational modes and optical phonons.
This localized probe of material stress is furthermore expected to preserve its large coupling strength when working with even smaller oscillators. 
Due to its high stiffness, semi-transparency behavior and, extremely low mass, MLG emerges as an ultra-sensitive platform for the simultaneous exploration of the spatial, temporal and spectral properties of NEMS, this system is thus promising for the detection of ultralow forces and could be used as carbon-based molecular sensors.
Moreover, this probe allows low temperature measurements, thus paving the way for stress mapping of other high quality factor resonators and understanding of dissipation factors in such systems.

\small
\acknowledgments

This work is partially supported by ANR grants (MolNanoSpin, Supergraph, Allucinan), ERC Advanced Grant  No. 226558 and, the Nanosciences Foundation of Grenoble. 
Samples were fabricated in the NANOFAB facility of the N\'eel Institute.
We thank A. Allain, D. Basko, C. Blanc, E. Bonet, O. Bourgeois, E. Collin, T. Crozes, L. Del-Rey, M. Deshmukh, E. Eyraud, C. Girit, R. Haettel, C. Hoarau, D. Jeguso, D. Lepoittevin, R. Maurand, J-F. Motte, R. Piquerel, Ph. Poncharal, V. Reita, A. Siria, C. Thirion, P. Vincent, R. Vincent and W. Wernsdorfer for help and discussions.

\bigskip

\textbf{Methods}
Multilayered Graphene flakes are deposited on 280 nm thick oxidized silicon wafer by micro-mechanical exfoliation\cite{Novoselov2004p2970} of Kish graphite.
Electrical contacts are made using deep UV lithography and e-beam deposition of  50 nm Au film.
Samples are suspended by etching (buffered hydrogen fluoride at concentration 1:3 HF/NH$_4$F) and drying using CO$_2$ critical point drying.
Experiments have been performed on approximately 100 samples (see Supp. Info).
Micro-Raman spectroscopy was performed with a commercial Witec Alpha 500 spectrometer setup with a dual axis XY piezo stage in a back-scattering/reflection configuration.
Grating used has 1800 lines/mm which confer a spectral resolution of 0.01 cm$^{-1}$ for 10 s integration time. 
Two laser excitation wavelengths are used, 633 nm (He-Ne) and 532 nm (Solid state argon diode).
Raman spectra are recorded in air with a Nikon x100 objective ($NA=0.9$) focusing the light on a 320 nm diameter spot (532 nm light) and, with a Mitutoyo x50 objective (M plan APO NIR) in vacuum.
All measurements made under vacuum (Fig. 3, Fig. 4a-b) are under active pumping at residual pressure equals to 10$^{-6}$ bar.
For Raman (reflectance) measurements laser power is kept below 1mW.$\mu$m$^{-2}$  (1 $\mu$W.$\mu$m$^{-2}$).
For rf measurements in air or vacuum, optical response is recorded with a silicon fast photodiode and a lock-in detector synchronized at twice the rf drive frequency.
The signal is maximum when 2$\ \omega_{AC}$ coincides with the fundamental mechanical resonance frequency $\omega_0$ of the cantilever.
Cryogenic measurements involve an optical continuous He flow Janis cryostat with electrical contacts.

%%%%%%%%%%%%%%%%%%%%%%%%%%%%%%%%%%%%%%%%%%%%%%%%%%%%%%%%%%%%%%%%%%%%%%%%%%%%%%%%%%%%%%%%%%%%%%%%%%%%%%%%%%%%%%%%%
%%%%%										BIBLIOGRAPHY   					      %%%%%%%%%%%%%%%%%%%%%%%%%%%%%%%%%%%%%%%%%%%%%%%%%%%%%%
%%%%%%%%%%%%%%%%%%%%%%%%%%%%%%%%%%%%%%%%%%%%%%%%%%%%%%%%%%%%%%%%%%%%%%%%%%%%%%%%%%%%%%%%%%%%%%%%%%%%%%%%%%%%%%%%%%%%%%%%%%%%%%%%%%%%%%%%%%%%%%%%%%%%%%%%%%%%%%%%%%%%%%%%%%%%%%%%%%%%%%%%%%%%%%%%%%%%%%%%%%%%%%%%%%%%%%%%%%%%%%

\end{document}